\documentclass[twoside,11pt]{article}

\usepackage{jmlr2e}
\usepackage{mathtools}
\usepackage{graphicx}
\usepackage{subfig}
\usepackage{float}
\usepackage{multirow}
\usepackage{nicematrix}
\usepackage{booktabs}
\usepackage{hyperref}
\usepackage[utf8]{inputenc}

\DeclarePairedDelimiter\floor{\lfloor}{\rfloor}

\firstpageno{1}

\begin{document}

\title{Short-Time Fourier Transform for deblurring Variational Autoencoders}

\author{\name Vibhu Dalal \\
       \medskip\addr Sri Aurobindo International Centre of Education\\
       Puducherry, India}

\maketitle

\begin{abstract}
Variational Autoencoders (VAEs) are powerful generative models, however their generated samples are known to suffer from a characteristic blurriness, as compared to the outputs of alternative generating techniques. Extensive research efforts have been made to tackle this problem, and several works have focused on modifying the reconstruction term of the evidence lower bound (ELBO). In particular, many have experimented with augmenting the reconstruction loss with losses in the frequency domain. Such loss functions usually employ the Fourier transform to explicitly penalise the lack of higher frequency components in the generated samples, which are responsible for sharp visual features. In this paper, we explore the aspects of previous such approaches which aren't well understood, and we propose an augmentation to the reconstruction term in response to them. Our reasoning leads us to use the short-time Fourier transform and to emphasise on local phase coherence between the input and output samples. We illustrate the potential of our proposed loss on the MNIST dataset by providing both qualitative and quantitative results. 
\end{abstract}


\section{Introduction}

\begin{figure}[H]
    \vspace{-3mm}
  \centering
  \subfloat[L2 loss]{\includegraphics[width=0.45\linewidth]{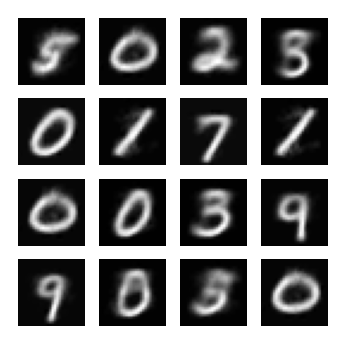}\label{fig:1}}\hspace{5mm}
  \subfloat[SSIM loss]{\includegraphics[width=0.45\linewidth]{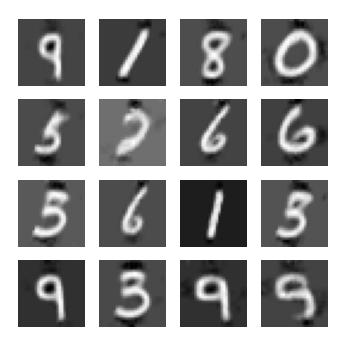}\label{fig:2}}\hspace{1em}
  \subfloat[DFT+SSIM]{\includegraphics[width=0.45\linewidth]{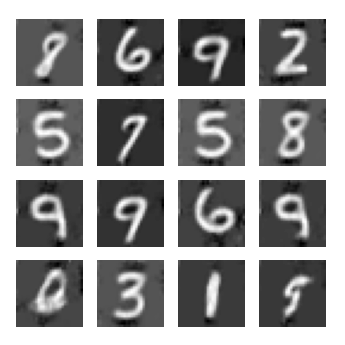}\label{fig:3}}\hspace{5mm}
  \subfloat[Ours]{\includegraphics[width=0.45\linewidth]{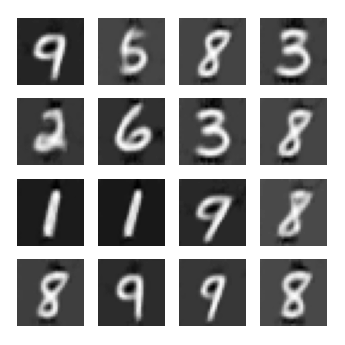}\label{fig:4}}\vspace{2mm}
  \caption{Samples generated after training the VAE on the MNIST dataset. In (a) we see the characteristic blurriness of VAEs by using MSE as the reconstruction loss term. In (b) we already see an improvement over (a) but the digits still possess a degree of roughness and blur. In (c) we notice that both blurriness and roughness have reduced slightly.}
  \label{results}
\end{figure}

Generative modelling is a broad area of machine learning that aims to learn a data distribution from a given set of samples, such that novel samples can be generated from the learned distribution. With the advent of deep learning, deep generative models, which are neural networks trained to approximate complicated, high-dimensional probability distributions, have soared in popularity (\cite{ruthotto2021introduction}, \cite{bond2021deep}). One such generative modelling approach of particular interest is Variational Bayes, which is employed by the Variational Autoencoder (VAE) as introduced by \cite{kingma2013auto} to learn a latent variable model. \\\\
VAEs belong to the families of probabilistic graphical models and have become increasingly popular due to their strong probabilistic foundations and latent variable model architecture. They have become one of the major areas of research in generative modelling, and have supplanted other autoencoder approaches. In terms of of architecture, they employ an encoder-decoder pair and use backpropagation to directly maximise the evidence lower bound (ELBO). The term ‘variational’ stems from the variational inference technique which is used to regularise the latent space. But despite their popularity and achievements, the behaviour of VAEs is still often far from satisfactory. Several theoretical and practical hindrances exist and represent active areas of research. The presence of blurriness in generated images is one such drawback of VAEs.\\\\
In comparison with other generative techniques, VAEs are known to produce images with a characteristic blurriness. As explained by \cite{bredell2022explicitly}, this is mainly due to the formulation of the optimisation. VAEs are optimised by maximising the evidence lower bound (ELBO) $\mathbb{E}_{z\sim q_{\phi}(z|x)}\: log[p_{\theta}(x|z)] - D_{KL}[q_{\phi}(z|x)||p(z)]$ with respect to the network parameters $\theta$ and $\phi$. This optimisation objective can equivalently be formulated as minimising the {\em Kullback-Leibler} divergence between the original data distribution $q_{D,\phi}(x,z)$ and $p_{\theta}(x,z)$. Now due to the asymmetry of the {\em Kullback-Leibler} divergence, the model will be penalised heavily if the samples are likely under $q_{D,\phi}(x,z)$ but not under $p_{\theta}(x,z)$. A consequence of this would be that $p_{\theta}(x,z)$ will have a larger variance than the original data distribution, leading to generated samples being more diverse and potentially more blurry. Several attempts have been made to resolve this issue by manipulating the reconstruction term, and one popular method is that of using a frequency loss function in the Fourier domain.\\\\
To devise a frequency loss function, the Fourier transforms of both the input and output images are first calculated and then a function is designed to penalise the differences in features between the two transforms. The loss function usually penalises the higher frequencies more since blurriness in an image is often associated with a lack of high-frequency components in the frequency spectrum of the image.\\\\ 
\cite{jiang2021focal} proposed a simple yet effective frequency loss function called the focal frequency loss. They first map each spectrum coordinate value to a Euclidean vector in two-dimensional space using both the amplitude and phase information. The proposed loss is then defined as the scaled Euclidean distances of the corresponding vectors of input and output images. As another instance, \cite{fuoli2021fourier} first compute the L1 distances between the frequency amplitudes and phases of the input and output images, and then proceed to evaluate the frequency loss which is the sum of the frequency amplitude and phase losses. Some other instances of using frequency losses include the works of \cite{bjork2022simpler} and \cite{bredell2022explicitly}. In all the aforementioned instances, the frequency loss functions are used in combination with pixel-wise loss functions. And importantly, they all base themselves on the global frequency spectrum of the images.\\\\
Inspired by the findings of \cite{wang2003local}, which illustrate that local frequency components in the Fourier domain can be more relevant in the preservation of image information than the global frequency components, we propose a novel frequency loss function for the Variational Autoencoder which bases itself on the local frequency amplitude and phase values of an image. The insights presented in \cite{article} and \cite{jiang2021focal} further help us to frame our loss function. To the best of our knowledge, ours is the only such work which applies a local frequency loss as a supervision loss for the variational autoencoder.\\\\
We use a small-scale VAE and test our proposed loss function on the MNIST dataset, and achieve a SSIM score of 0.492 and a PSNR score of 11.93, and show that our method's performance marginally beats the performances of alternate loss functions. We also provide a qualitative analysis of the results in \ref{results}.
\section{Background}
\begin{figure}[H]
    \centering
    \includegraphics[width=0.75\linewidth]{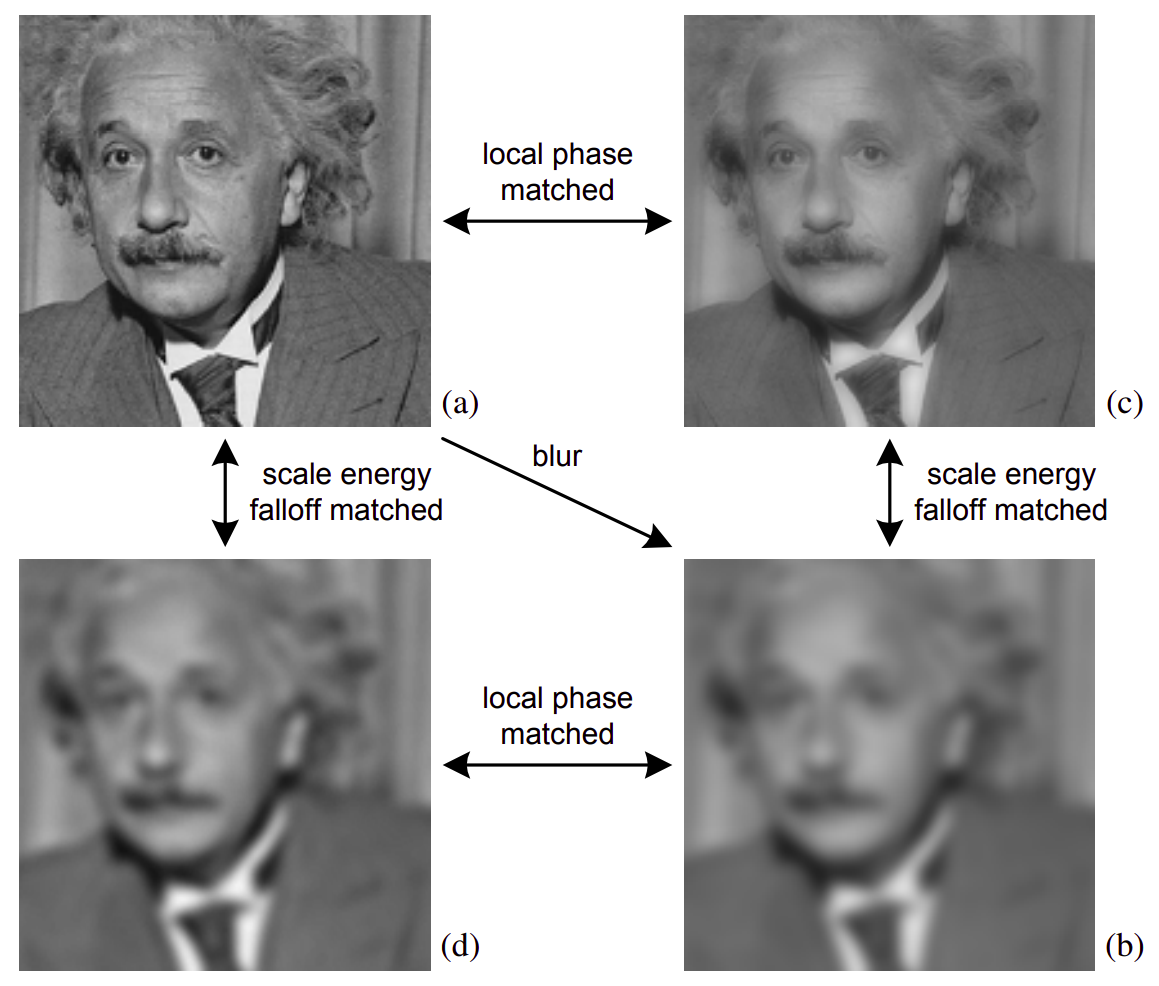}
    \caption{Illustration of the significance of local phase in blur perception. (a) original image; (b) blurred image obtained by convolving with a circular-symmetric Gaussian low-pass filter; (c) image (a) with high-frequency band energy reduced to match that of (b)'s; (d) image (b) with high-frequency band energy elevated to match that of (a)'s.}
    \label{fig:localphase}
\end{figure}

As explained in \cite{fuoli2021fourier}, the main advantage of using a frequency loss function is that it explicitly penalises the difference between the high-frequency components of input and output images, which promotes the retention of sharp features in the generated image. This is opposed to pixel-wise loss functions such as the $L1$ or $L2$ loss functions which are known to be insensitive to blurriness (\cite{bredell2022explicitly}). However, as discussed in the approaches mentioned in the previous section, using a frequency loss function in combination with a pixel-wise loss function is empirically shown to produce the best results.\\\\
Previous works that have used frequency loss functions in the training of Variational Autoencoders employed the global Fourier transform of the input and output images, which is given by:
$$F(u,v) = \frac{1}{\sqrt{HW}} \sum_{h=0}^{H-1}\sum_{w=0}^{W-1} f(h,w) e^{-i2\pi\left(\frac{uh}{H}+\frac{vw}{W}\right)} $$\\
where the input image is of size $H \times W$, and $f(h,w)$ is the value of the pixel at $(h,w)$. 
\noindent As seen in the previously mentioned works, such frequency loss functions involve the use of the $L1$ or $L2$ functions to compute some form of dissimilarity between the global Fourier transforms of the input and output images. As explored in \cite{jiang2021focal}, a weighting scheme is sometimes used to accentuate the differences in the higher frequency values of the images, which corresponds to penalising the output image for not possessing the sharper features of the input image. However, there are several details regarding the framing of these loss functions which are not well understood.\\\\
Firstly, there is no clear insight as to which component of the global Fourier transform -- amplitude or phase -- should be emphasised more in the formulation of the frequency loss function. The aforementioned approaches seem to provide equal emphasis to both components, which might not necessarily be the optimal strategy. Secondly, the advantage of the global Fourier transform over the local or short-time Fourier transform is not well understood, especially when it pertains to blurriness. These details might be filled by examining the ideas explored in the works of \cite{murray2010perceived} and \cite{wang2003local}.\\\\
\cite{murray2010perceived} had previously questioned the traditional view that perceived blur in an image is due to reduced energy at high frequencies, and had emphasised the importance of the phase of high spatial frequencies. Meanwhile, \cite{wang2003local} argue that the distortions of local phase contribute more to the perception of blurriness than the widely noted loss of high-frequency energy. They also illustrate this qualitatively by showing that a sharp image with its high frequency energy reduced but local phase preserved appears much sharper than a blurred image with its high frequency energy corrected but local phase uncorrected, as shown in \ref{fig:localphase}. In the context of training Variational Autoencoders, these findings would suggest that it might be advantageous to use the short-time Fourier transform and accentuate the differences in the local phase values between the input and output images in order to emphasise local phase coherence.
\vspace{2mm}

\section{Methodology}
\subsection{Frequency loss function}
A $\beta$ Variational Autoencoder is optimised by maximising the evidence lower bound (ELBO) which is given by:
$$L(q) = \mathbb{E}_{z\sim q_{\phi}(z|x)}\: log[p_{\theta}(x|z)] - \beta D_{KL}[q_{\phi}(z|x)||p(z)]$$ 

\noindent The first term is referred to as the reconstruction loss which ensures that a sample can be reconstructed from its latent representation, and the second term is the {\em Kulbeck-Leibler} divergence between the learned approximate posterior distribution and the prior distribution $p(z)$ which is usually $\mathcal{N}(0,1)$.\\\\
In order to explicitly minimise blur in the generated samples, we formulate the reconstruction loss as a combination of a frequency loss and a pixel-wise loss. Inspired by the findings discussed in the previous section, we formulate our frequency loss function as follows.\\\\
Let $F(u,v,m_0,n_0)$ be the short-time Fourier transform of an $M\times N$ image, which is given by:
$$F(u,v,m_0,n_0) = \sum_{m=0}^{M-1}\sum_{n=0}^{N-1} f(m,n)\,\mathcal{W}(m-m_0,n-n_0) e^{-i[u(m-m_0)+v(n-n_0)]}$$\\
where $\mathcal{W}(x,y)$ is a real valued windowed function centered at $(x,y)$ which is used to avoid edge induced artifacts. Inspired by \cite{fuoli2021fourier}, we employ an $L \times L$ Hann window as the windowed function, which is given by:
$$\mathcal{W}(x,y) = \frac{1}{4}\left(1-\cos{\frac{2\pi x}{L}}\right)\left(1-\cos{\frac{2\pi y}{L}}\right), \; 0\le x,y<L$$ 
\\
and $0$ otherwise. Thus computing the short-time Fourier transform of an image amounts to sliding the Hann window over all possible $(m_0,n_0)$ pairs and then computing the discrete Fourier transform of the resulting image. If $h_l$ and $h_s$ are the side length and stride of the Hann window respectively, then the output of the short-time Fourier transform algorithm performed on an $M\times N$ image is a $K\times L\times h_l\times h_l$ matrix where
$$K=\floor*{\frac{M-h_l}{h_s}}+1, \hspace{4mm}
L=\floor*{\frac{N-h_l}{h_s}}+1$$\\
\noindent Let $F_i$ and $F_o$ denote the short-time Fourier transforms of the input and output images, and let $A_i$, $P_i$ and $A_o$, $P_o$ denote their respective amplitude and phase components. The matrices $A_i$ and $A_o$ contain the local amplitudes of the input and output images respectively, and the matrices $P_i$ and $P_o$ contain their local phases. They are given by:
$$A_{i_{h,w}} = |F_{i_{h,w}}| = \sqrt{R(F_{i_{h,w}})^2+I(F_{i_{h,w}})^2}$$
$$P_{i_{h,w}} = \angle{F_{i_{h,w}}} = atan2(I(F_{i_{h,w}}), R(F_{i_{h,w}}))$$\\
Next, we use the $L1$ function to compute the dissimilarities between $A_i$ and $A_o$, and $P_i$ and $P_o$. Moreover, we use a linearly increasing weighting scheme to emphasise the local amplitudes and phases pertaining to higher frequencies, in order to explicitly tackle the blurriness in the generated image. Additionally, inspired by the findings of \cite{wang2003local}, we apply a higher relative weighting to the local phase loss by multiplying it with a tunable hyperparametre $\lambda$. The resulting frequency loss function becomes: 
$$L_{freq} = \sum_{j}(\lambda|P_{o_j}-P_{i_j}| + |A_{o_j}-A_{i_j}|)W_{j}, \;\lambda > 1$$\\
where $W_j$ is the corresponding weighting term meant to accentuate the losses of higher frequencies.
\subsection{Pixel-wise loss function}
We experiment with a variety of pixel-wise loss functions such as the $L1$, $L2$, sigmoid cross entropy and the SSIM loss functions, and we find that the latter seems to work best with our proposed frequency loss function. The Structural Similarity Index (SSIM) is a metric for perceptual difference that compares two images on the bases of three features: luminance, contrast and structure. The SSIM between two images is a value between -1 and 1, where the former signifies that the given images are very different, and the latter signifies that the images are very similar or the same. The reader may refer to \cite{gilbertnon}, \cite{zhao2016loss} for more information on the SSIM.\\

\noindent The resulting reconstruction loss between the input sample $S_i$ and the output sample $S_o$ can now be written as
$$L_{recons} = \lambda_{freq}L_{freq} + SSIM(S_i, S_o)$$where $\lambda_{freq}$ is a tunable hyperparametre.\\\\
Finally, the complete loss function of the Variational Autoencoder can be written as:

\begin{align}
L &= \beta D_{KL}[q_{\phi}(z|x)||p(z) + L_{recons} \\
& = \beta D_{KL}[q_{\phi}(z|x)||p(z) + \lambda_{freq}L_{freq} + SSIM(S_i, S_o)
\end{align}

\section{Experiments and Results}
To evaluate our model on a benchmark dataset, we used a fixed binarized version of the MNIST digit dataset defined by Larochelle and Murray [2011]. It consists of 60,000 training and 10,000 test images of handwritten digits (0 to 9) which are 28 × 28 pixels in size. The code is written in Python and Tensorflow is used for training the models. We use a simple VAE architecture similar to the one described in \cite{tfvae}, which consists of 2 convolutional encoding and decoding layers that are connected through a multilayer perceptron which does the final compression to the latent variable size, which is 2, and the initial scaling up thereafter. We choose a compact architecture due to our hardware limitations.
Details on the architecture can be found in the code that is publicly available on Github.\footnote{\url{https://github.com/Vibhu04/Deblurring-Variational-Autoencoders-with-STFT}}\\\\
Furthermore, the Adam optimizer is used with a linearly decreasing cyclical learning rate as described in \cite{smith2017cyclical}. We set the minimum and maximum learning rates to $10^{-4}$ and $10^{-3}$ respectively, and the step size to 2000. We train for 50 epochs with a batch size of 50. Additionally, we experiment with a range of short-time Fourier transform window dimensions and strides, and find that the best results are achieved with a 16 x 16 window and a stride of 4.\\\\
We used the peak-signal to noise ratio (PSNR) and the structural similarity measure index (SSIM) as metrics for comparing the reconstruction with its corresponding input sample. We provide quantitative and qualitative assessments of 3 reconstruction loss functions: 1) mean squared error (MSE), 2) SSIM, and 3) the full frequency loss function defined above.

\begin{table}
    \centering
    \begin{NiceTabular}{ccc}
        \toprule
        Reconstruction loss function & PSNR & SSIM\\
         \midrule
         L2 & 10.20 & 0.41\\
         L1 & 10.13 & 0.32\\
         SSIM & 11.38 & 0.495\\
         DFT+SSIM & 11.51 & 0.486\\
         \RowStyle{\bfseries}Ours & 11.93 & 0.492\\
         \bottomrule
    \end{NiceTabular}
    \caption{We provide quantitative results on the MNIST dataset. It can be seen that our proposed loss function marginally outperforms the SSIM loss function, which is actually one of its components.}
\end{table}

\section{Conclusion}
In this paper, we have proposed a reconstruction loss function which employs a frequency loss function in the Fourier domain and the SSIM function to minimise the characteristic blurriness in the generated samples of a VAE. We differentiate our approach from others by using the short-time Fourier transform instead of the global Fourier transform, and by emphasising more on local phase coherence than local amplitude differences. We also weigh the differences in higher frequency components between the input and the corresponding output more in order to explicitly encourage the generated samples to have sharper features. We perform experiments on the MNIST dataset to compare the results of our proposed reconstruction loss with the results of other popular reconstruction losses, and we provide both qualitative and quantitative evaluations.

\bibliography{ref}

\end{document}